\documentclass[useAMS,usenatbib,usegraphicx]{mn2e}

\title[Southern ultracool dwarfs]{New nearby, bright southern ultracool 
dwarfs\footnotemark[0]\thanks{Based on observations made at
the European Southern Observatory, Chile (NTT/SOFI programs 076C.0382, 076D.0872 and 077C.0117)
and on data from the 2MASS project (University of Massachusetts and IPAC/Caltech, USA).}\\}
\author[T.R. Kendall et al.]
{T.R. Kendall$^{1}$\thanks{E-mail:
trk@star.herts.ac.uk}, H.R.A. Jones$^1$, D.J. Pinfield$^1$, R. S. Pokorny$^2$, S. Folkes$^1$,
\vspace*{2mm}\\
\vspace*{2mm}{\rm\LARGE D. Weights$^1$, J.S. Jenkins$^1$ and N. Mauron$^3$}\\
$^{1}$Centre for Astrophysics Research, 
Science and Technology Research Institute,
University of Hertfordshire, College Lane,\\Hatfield AL10 9AB, United Kingdom\\
$^{2}$Yunnan Observatory Chinese Academy of Sciences, PO Box 110, 650011, Kumming, China\\
$^{3}$Groupe d'Astrophysique, UMR 5024 CNRS, Case CC72, Place Bataillon, 34095 Montpellier Cedex 5, France}
\begin{document}

\date{Accepted; Received; in original form}

\pagerange{\pageref{firstpage}--\pageref{lastpage}} \pubyear{2006}

\maketitle

\label{firstpage}

\begin{abstract}
We report the discovery of twenty-one hitherto unknown bright southern
ultracool dwarfs with spectral types in the range M7 to L5.5, together 
with new observations of a further three late M dwarfs previously confirmed.
Three more objects are already identified in the literature as high proper motion stars;
we derive their spectral types for the first time. All
objects were selected from the 2MASS All Sky and SuperCOSMOS point source
databases on the basis of their optical/near-infrared colours, $J$-band magnitudes and 
proper motions. Low resolution (R\,$\sim$\,1000) $JH$ spectroscopy
with the ESO/NTT SOFI spectrograph has confirmed the ultracool nature of 
24 targets, out of a total of 25 candidates observed. Spectral types are derived 
by direct comparison with template objects and compared to results from 
H$_2$O and FeH indices. We also report the discovery of one binary, as revealed by 
SOFI acquisition imaging; spectra were taken for both components. The spectral types of the
two components are L2 and L4 and the distance $\sim$\,19\,pc.
Spectroscopic distances and transverse velocities are derived for the
sample. Two $\sim$\,L5 objects lie only $\sim$\,10\,pc distant. 
Such nearby objects are excellent targets for further study
to derive their parallaxes and to search for fainter, later companions
with AO and/or methane imaging. 
\end{abstract}


\section{Introduction}

In recent years, the search for faint dwarfs in the Solar neighbourhood has been the subject of much attention.
Such studies have been motivated by our still highly incomplete knowledge of the Sun's coolest and least massive
neighbours, as pointed out by, for example, \citet{hen97}. M dwarfs account for around 70\% of the stellar number density
in the Solar neighbourhood, and yet \citet{hen02} estimated as many as 63\% of stellar systems within 25\,pc remain
undiscovered, while the figure for missing systems within 10\,pc may be near 25\% \citep{rei03}, or even higher 
(30\%; \citealt{hen02}). Current and recent searches rely primarily on proper motion (\citealt{sub05}; \citealt{dea05}; \citealt{pok03}), 
in conjunction with photometry and spectroscopic follow-up (\citealt{hen04}; \citealt{lod05}; \citealt{sch05}; \citealt{pha06}; \citealt{pok05}).   

Efforts in the field have been enormously aided by
the digitization of wide-field survey plates by e.g. the SuperCOSMOS project (\citealt{ham01}; \citealt{hen04}), and by
the advent of all-sky near-infrared surveys, most notably 2MASS \citep{skr97} and DENIS \citep{epc97}, together with the Sloan Digital Sky Survey 
(SDSS; \citealt{yor00}). 

Moreover, observations of the low-mass dwarf population in the Solar neighbourhood have required the establishment of two new
spectral classes; the L and T dwarfs (\citealt{kir99}; \citealt{mar99}), with effective temperatures below that of the coolest M dwarfs.
These discoveries have resulted in a large part from the exploitation of all-sky surveys: we may cite \citet{bur04a} and \cite{chi06}
as two recent examples notable in for finding large numbers of both L and T dwarfs.
In the near future, the field will be further boosted by next-generation infrared surveys, notably the UKIDSS Large Area Survey
\citep{law06}. New L and T dwarf discoveries are increasingly populating the immediate Solar neighbourhood.

In this paper, we report the discovery of twenty-one new ultracool dwarfs (later than M6; \citealt{bou05}), the majority of which are estimated spectroscopically to lie
closer than $\sim$\,30\,pc, and which have spectral types later than M7. The two latest objects, with $\sim$\,L4-L6 spectral types, are 
good candidates to be very nearby; close to 10\,pc, and with reference to theoretical isochrones \citep{bar03}, are very likely substellar, i.e.
with masses below the hydrogen burning limit of 0.072\,M$_{\odot}$ \citep{leg01}. In the next Section, we overview the aims of this research and
methods of target selection. The spectroscopic observations are detailed in Sect. 3. In Sect 4., 
we present our results and discuss the methods we have used to analyse our spectra and derive spectroscopic distances and transverse
velocities for the targets. These results are discussed in Sect. 5. In Sect. 6 we show that one of the targets is a binary, 
with $\sim$\,L2 and $\sim$\,L4 components. A summary of our conclusions is presented in Sect. 7.

\begin{table*}
\vbox to 220mm{\vfil Landscape Table 1 to go here.  
\caption{} \vfil} \label{t1}
\end{table*}

\section[]{Survey aims and target selection}

\subsection{Survey overview and goals}

The aim of this project is to locate the remaining undiscovered, yet apparently bright, southern ultracool dwarfs in the Solar neighbourhood. 
By primarily using a
combination of their red optical and near-infrared colours, and secondarily {\it via} selecting objects from this
photometric sample which have the highest reduced proper motions, 
the survey presented here is designed to reveal bright, and therefore nearby dwarfs undiscovered by previous studies.
By requiring candidates to have an $R$-band detection, the search is biased towards nearer and/or more intrinsically
bright, earlier type dwarfs. A secondary bias is introduced
by the use of the reduced proper motion, which as a combination of absolute magnitude and transverse velocity \citep{pok03} may lead to
the preferential discovery of high velocity objects with possible thick-disk or halo kinematics. In general, however, our aim is to find those 
ultracool dwarfs, optically bright enough to be detected in the SuperCOSMOS database, which lie above our reduced proper motion threshold. This
corresponds to only a few $\times$\,10$^2$\,mas\,yr$^{-1}$ in most cases (see Table \ref{t1}).

\subsection{Target selection}

\begin{figure}
\includegraphics[width=9cm]{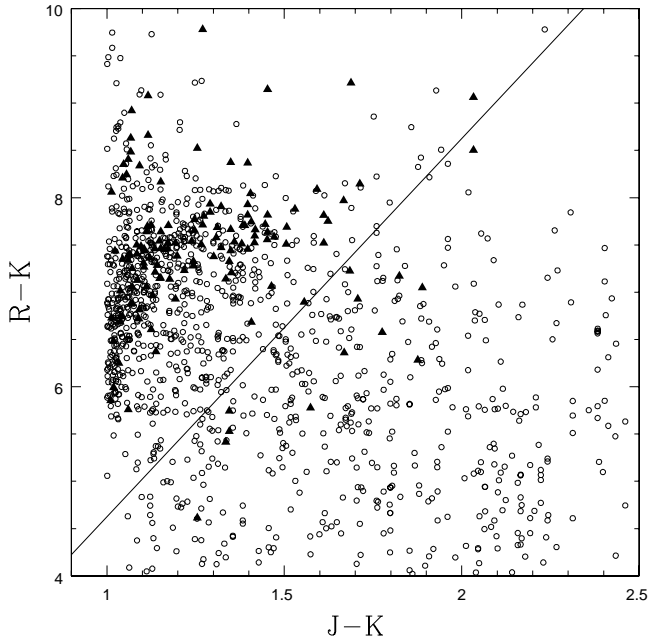}
 \caption{Illustration of the $R-K, J-K$ cut used to select candidate ultracool dwarfs. Known such objects (filled triangles) lie above and to the left of
the line shown. Candidates (circles) in this region were retained.}
\label{z1}
\end{figure}

\begin{figure}
\includegraphics[width=9cm]{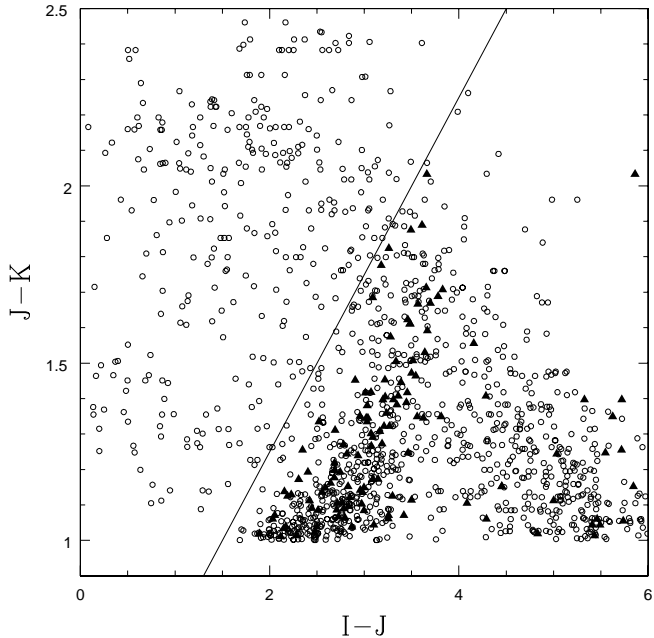}
 \caption{As Fig.\,\ref{z1}, for $J-K, I-J$. Candidates lie below and to the right of the line shown. }
\label{z2}
\end{figure}

\begin{figure}
\includegraphics[width=9cm]{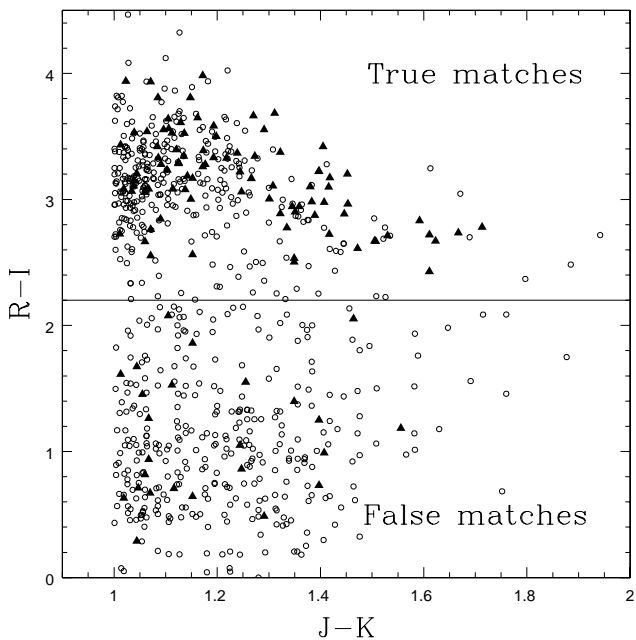}
 \caption{Illustration of the removal of false matches with the SuperCOSMOS $R-I$ colour. True matches to known ultracool
dwarfs lie above the horizontal line. }
\label{z3}
\end{figure}

\begin{figure}
\includegraphics[width=9cm]{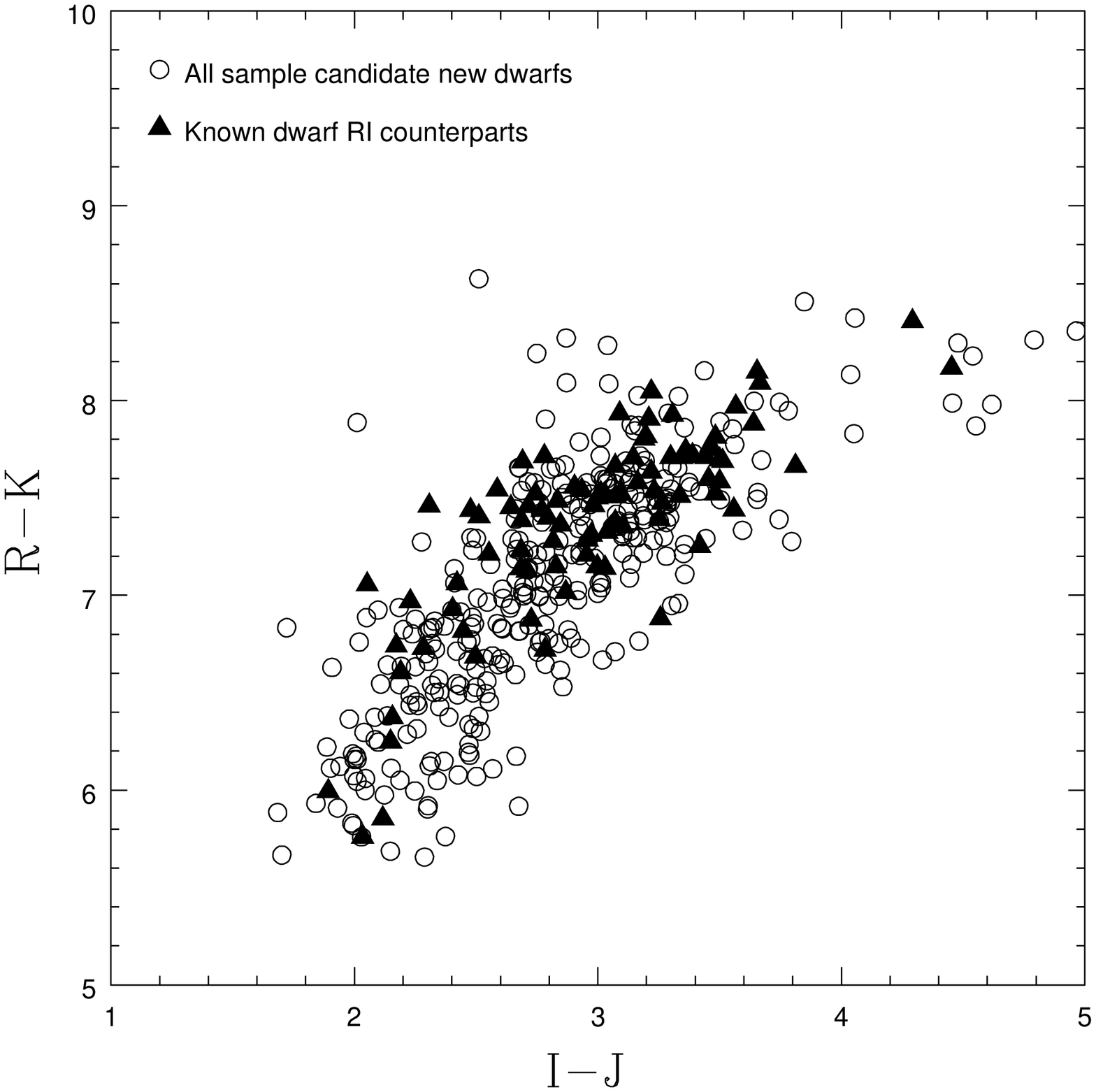}
 \caption{Efficiency of the $R-K$ colour to select ultracool dwarfs. The plot shows 90 out of 103 known dwarfs with spectral types 
M6 or later and with SuperCOSMOS $RI$ counterparts retained by our photometric selection methods (triangles). Circles are 317
candidates occupying the same region of the $R-K,I-J$ diagram, from which 61 high proper motion candidates were drawn.}
\label{f1}
\end{figure}

Targets were selected by interrogation of the current 
photometric and kinematic data provided by the 2MASS All-Sky point source and SuperCOSMOS\footnote{{\tt http://surveys.roe.ac.uk/ssa/}} 
surveys covering the complete Southern Hemisphere. Candidate
ultracool dwarfs were drawn from 2MASS with $|b|$\,$>$\,10$^{\circ}$, and with $JHK$ data passing cuts based on those of 
\citet{cru03}. Preliminary cuts on the 2MASS sample were:

\begin{description}
\item{11\,$\leq$\,$J$\,$\leq$\,17

\item $J-K$\,$>$\,1

\item $J$\,$\leq$\,3\,$\times$\,$(J-K)$\,+\,10.5}
\end{description}

\noindent yielding an initial sample of 1055887 2MASS objects, which was immediately reduced to 109237
by removing any point with photometric quality flag 'U' (upper limit) on any magnitudes.

Excluding 52387 objects in regions of high stellar surface density (e.g. the LMC and SMC, star-forming and other regions at low galactic
latitude defined by \citet{cru03}, with very slight modifications), the search yielded
8101 objects simultaneously satisfying the following cuts: 

\begin{description}
\item{$(J-H)$\,$\leq$\,1.75\,$\times$\,$(H-K)$\,+\,0.1875

\item$(J-H)$\,$\geq$\,1.75\,$\times$\,$(H-K)$\,-\,0.475}
\end{description}

and additionally:

\begin{description}
\item{$(J-H)$\,$\leq$\,0.8 where 0.3\,$\leq$\,$(H-K)$\,$\leq$\,0.35 

\item $(J-H)$\,$>$\,1\,--\,$(H-K)$}
\end{description}

\noindent where the last cut is a translation of the initial $J-K$\,$>$\,1 criterion.

Objects with $\delta$\,$\ga$\,3$^{\circ}$ were not considered further, so as to ensure SuperCOSMOS optical coverage of the
sample,  keeping 4074 objects, and of these only 2MASS 
photometry better than flag\,=\,CCC (SNR\,$\geq$\,5 and $\Delta$$JHK$\,$\leq$\,0.2), was kept, retaining 2317 objects.

A search radius of
18\arcsec\,was used to search for a SuperCOSMOS optical counterpart for each object, 
to allow for proper motion, and which,  given the typical difference in epoch between the two surveys 
($\sim$\,20\,yr) allowed retention of candidate field dwarfs with $\mu_{\rm tot}$ up to $\sim$\,1\,\arcsec\,yr$^{-1}$. 
Although closer epoch differences could allow higher $\mu$ objects through, with possibly less accurate measurements,
this survey has not been designed to yield the very highest proper motion objects, which are in any case expected to 
be far less numerous. In contrast to, for example \citet{pok05} and \citet{dea05}, this research instead is
primarily designed to find late-type ($\ga$\,M7), ultracool dwarfs {\it via} a combination of their optical and
near-infrared colours, using proper motion to remove more distant contaminants. 

At the same time, a total of 135 known ultracool objects taken mostly
from \citet{cru03} were found in the SuperCOSMOS data (i.e. there was one or more optical counterpart with $R$ and $I$
photometry) and a  total of 1245 other counterparts to the 2MASS data were also found, but these included false matches and a high
proportion of contaminating objects (e.g. AGB carbon stars). Further cuts were therefore applied to the dataset of optical
counterparts to the 2317 2MASS sources, to retain as many known ultracool dwarfs
and discard other data points, using the loci of known objects in the $R-K,J-K$ (Fig.\,\ref{z1}) and $J-K,I-J$ (Fig.\,\ref{z2}) diagrams as a guide. These cuts were:

\begin{description}

\item{$(R-K)$\,$\geq$\,4$\times$\,$(J-K)$\,+\,0.625

\item $(J-K)$\,$\leq$\,0.5$\times$\,$(I-J)$\,+\,0.25}

\end{description}

together with:

\begin{description}
\item{5.5\,$\leq$\,$(R-K)$\,$\leq$\,9} 
\end{description}

\noindent to isolate L dwarfs.

\begin{figure}
\includegraphics[width=9cm]{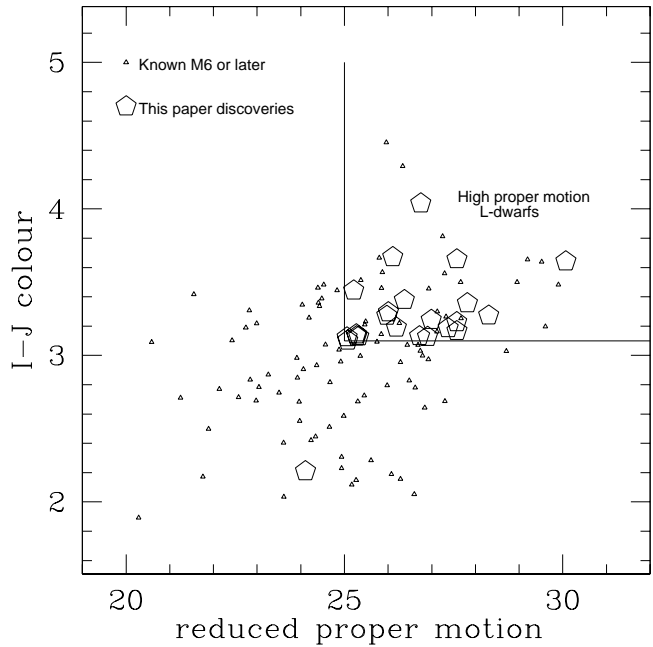}
 \caption{Selection of targets by reduced proper motion (H$_{\rm J}$\,=\,$J$\,+\,5log($\mu_{\rm tot}$) with $\mu$ in mas\,yr$^{-1}$)
and $I-J$ colour. Small triangles are data for the 90 known L dwarfs retained by our selection methods (see text) {\it after}
use of optical vs. infrared colour criteria. The 
new object outside the selection criteria was originally mismatched in the 2MASS/SuperCOSMOS cross-correlation.}
\label{f2}
\end{figure}

Moreover, false matches were almost completely removed by rejecting points with $(R-I)$\,$\leq$\,2.2 (Fig.\,\ref{z3}), although note that the SuperCosmos
$R$ filter does not have the same characteristics as Cousins $R$, so this last cut is not necessarily applicable when comparing 2MASS magnitudes to
optical magnitudes from other surveys. Ninety known dwarfs were 
retained out of 103 passing the first three near infrared-only colour cuts. 
New candidate L dwarfs with optical counterparts passing the above optical/near-infrared cuts numbered 317 in total. The efficiency of the 
$R-K$ colour to select ultracool dwarfs
is shown in Fig.\,\ref{f1} where the 90 known objects and 317 potential candidate new dwarfs are shown to occupy
the same region of the $R-K,I-J$ diagram. Indeed, a 2D Kolmogorov-Smirnov test shows this is the case with
99.3\% probability. Our final southern sample consists of 61 of these 317 , selected to have reduced
proper motion H$_{\rm J}$\,$\geq$\,25 (in milliarcsecond units) and $(I-J)$\,$\geq$\,3.1, typical for L dwarfs \citep{ken04}.
The 90 known L dwarfs passing these photometric cuts are plotted in Fig.\,\ref{f2} together with the 21 new discoveries 
presented here, and three further late M/early L objects from \citet{lod05} for which
we present new spectroscopic data. In the last part of the selection process, candidates 
appearing in the SIMBAD database, or as unpublished objects listed in DwarfArchives.org\footnote{{\tt www.dwarfarchives.org}},
or rejected by visual inspection of the SuperCOSMOS $R$-band images, 
were removed. 25 targets remained after this last rejection process, and 24 out of 25 targets observed spectroscopically
are found here to be ultracool dwarfs later than $\sim$\,M7, the one interloper being a reddened early-type object, as evidenced by the low stellar 
density in the surrounding 5\arcmin$\times$5\arcmin~ SuperCOSMOS $I$-band field and despite the high galactic latitude $b$\,=\,$-$36$^{\circ}$.
Optical and infrared photometry are given in Table \ref{t1} together with proper motions and other basic data. Derived spectral types, spectroscopic
distances and transverse velocities are given in Table \ref{t2}. We give details of the methods used to determine these in Sect. 4.

\section[]{Spectroscopic observations and data reduction}

Observations were carried out in three runs at the ESO New Technology Telescope (NTT) with the SOFI
spectrograph, during October 25--29, 2005, January 17--19, 2006 and April 4--7, 2006, allowing observation
of the complete sample of southern targets at all RAs. The 0.6\arcsec~ slit was employed with the blue $JH$
grism yielding a wavelength coverage of 9300--16500\,\AA~ at a spectral resolution R\,=\,$\lambda/\Delta\lambda$\,$\sim$\,1000.
The detector is a Hawaii HgCdTe 1024$\times$1024 array with 18.5\,\micron~pixels.
Accurate sky subtraction was facilitated by nodding the telescope along the slit, typically by 30--45\arcsec.
Detector integration times (DIT) were in the range 60\,sec to 120\,sec, with between 4 and 12 integrations
at each nod position, yielding total integration times for science targets ranging between 16 and 48\,min, depending on
target $J$ magnitude. The resulting spectra have signal-to-noise ratios of at least $\sim$\,100 and often better. 
Conditions were generally good with optical seeing in the range 
0.6\arcsec -- 1\arcsec; however during the January 2006 run in particular, conditions of high humidity saturated the
telluric H$_2$O band at 1.35\,\micron~ and prevented its accurate removal. A, F or late B-type 
telluric standards were observed before and after each target observation at airmasses within 0.1 and often within 
0.05 of the target. Dome flats and xenon lamp
arcs were taken each afternoon, and again at the end of the night. Observations of template M6.5, M9, L1, L3 and L5 dwarfs
were taken with the same instrumental setup; multiple observations of some of these standard stars were obtained
during different observing runs.

\begin{figure*}
\includegraphics[bb=0 0 543 700,width=17cm]{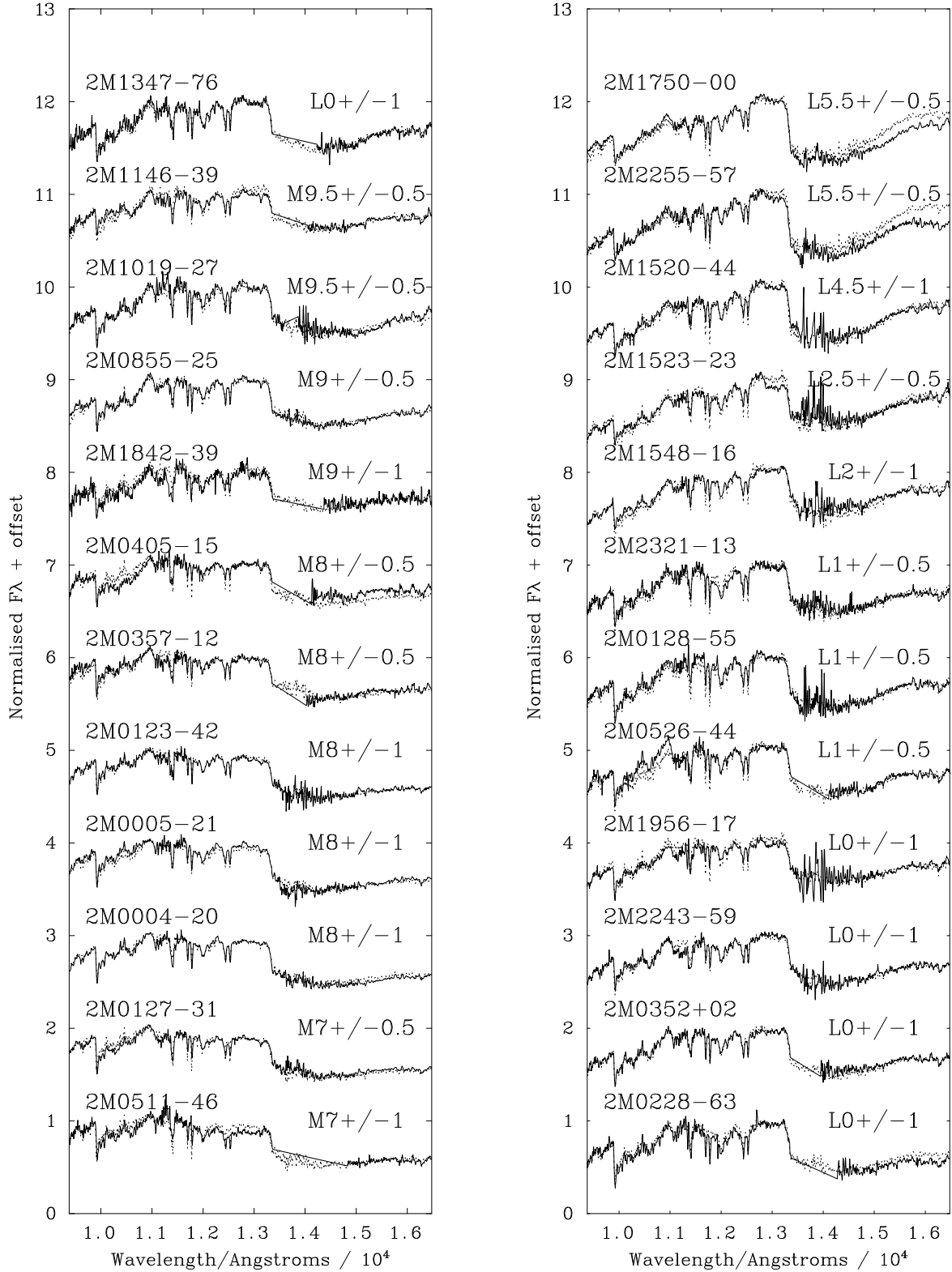}
 \caption{Spectral types by direct comparison to SpeX standards \citep{cus05} and known objects observed with the 
same instrumental setup as the targets. Unweighted means of adjacent
($\la$\,1 subclass) standards from both datasets were permitted to obtain the best fits to the
spectra for new objects.}
\label{f3}
\end{figure*}

\begin{table*}
\centering
 \begin{minipage}{110mm}
  \caption{Derived data for the target objects. Comparison of spectral types from direct comparison (col. 5, see Fig.\,\ref{f3}) 
with spectral types derived from the spectral indices of
\citet{sle04} (cols. 2 \& 4) and \citet{mcl03} (col. 3; see Sect. 4.2). 
Where the H$_2$O-l index is affected by poor telluric correction, no value is given. 
The FeH index is invalid for spectral types later than $\sim$\,L3. Col. 6 gives the spectroscopic distance with errors reflecting
the stated error on the spectral type, itself derived by comparison to template objects. Errors on transverse velocities (Col. 7)
reflect only the error on the distance. }
  \begin{tabular}{@{}lllllll@{}}
  \hline
Name & H$_2$O-l & H$_2$OB & FeH & spec. & d/pc & v$_{\rm trans}$/km\,s$^{-1}$ \\
\hline
2M0004$-$20 & L0.5 & L0 & M9 & M8\,$\pm$\,1 & 17.7$^{+3.8}_{-2.4}$ &       69.3\,$\pm$\,12.1\\	  
2M0005$-$21 & L0 & M9.5 & M9 & M8\,$\pm$\,1 & 26.5$^{+5.7}_{-3.5}$ &    	83.0\,$\pm$\,14.4\\	 
2M0123$-$42 & M8 & M9.5 & M9 & M8\,$\pm$\,1 & 25.1$^{+5.4}_{-3.3}$ &       31.8\,$\pm$\,5.4\\	 
2M0127$-$31 & M9 & M9 & M7 & M7\,$\pm$\,0.5 & 24.3$^{+3.3}_{-2.4}$ &    	39.4\,$\pm$\,4.5\\	 
2M0128$-$55 & L2.5 & L3 & L1 & L1\,$\pm$\,0.5 & 22.7$^{+1.5}_{-1.5}$ &   	31.5\,$\pm$\,2.1\\	 
2M0228$-$63 & - & L1.5 & L2 & L0\,$\pm$\,1 & 23.2$^{+3.0}_{-2.7}$ & 	70.0\,$\pm$\,8.5\\	 
2M0352+02   & - & L0.5 & M9.5& L0\,$\pm$\,1 & 18.6$^{+2.4}_{-2.2}$ &   	40.2\,$\pm$\,5.0\\	 
2M0357$-$12 & - & M9 & M7.5 & M8\,$\pm$\,0.5 & 28.9$^{+2.7}_{-2.1}$ &  	18.5\,$\pm$\,1.5\\	 
2M0405$-$15 & - & M8.5 & M8 & M8\,$\pm$\,0.5 & 30.1$^{+2.8}_{-2.2}$ &  	40.1\,$\pm$\,3.3\\	 
2M0511$-$46 & - & M8.5 & M8 & M7\,$\pm$\,1 & 42.8$^{+14.0}_{-7.5}$  &  	35.1\,$\pm$\,8.8\\	 
2M0526$-$44 & - & L2 & M9.5 & L1\,$\pm$\,0.5 & 26.1$^{+1.7}_{-1.7}$ &  	20.8\,$\pm$\,1.4\\	 
2M0855$-$25 & L0 & L1 & M9.5 & M9\,$\pm$\,0.5 & 22.8$^{+1.6}_{-1.4}$ &   	41.8\,$\pm$\,2.8\\	 
2M1019$-$27 & M9 & L1 & L0 & M9.5\,$\pm$\,0.5 & 24.3$^{+1.6}_{-1.4}$ &	74.2\,$\pm$\,4.6\\	 
2M1146$-$39 & - & L1.5 & M9& M9.5\,$\pm$\,0.5 & 25.6$^{+1.6}_{-1.5}$ &	35.4\,$\pm$\,2.1\\	 
2M1347$-$76 & - & L2 & L1  & L0\,$\pm$\,1 & 25.9$^{+3.3}_{-3.0}$ &	47.3\,$\pm$\,5.7\\	 
2M1520$-$44 & L3 & L4 & - & L4.5\,$\pm$\,1 & 9.6$^{+2.2}_{-1.9}$ &    	33.7\,$\pm$\,7.0\\	 
2M1523$-$23 & L0 & L2.5 & L1 & L2.5\,$\pm$\,0.5 & 22.1$^{+1.8}_{-1.8}$ &	36.3\,$\pm$\,3.0\\	 
2M1548$-$16 & M8.5 & L1.5 & L0.5 & L2\,$\pm$\,1 & 20.8$^{+3.2}_{-3.2}$ &	18.6\,$\pm$\,2.9\\	 
2M1750$-$00 & L8 & L6 & - & L5.5\,$\pm$\,0.5 & 8.0$^{+0.9}_{-0.8}$ &    18.6\,$\pm$\,1.9\\	 
2M1842$-$39 & - & M6 & M9  & M9\,$\pm$\,1 & 29.7$^{+4.6}_{-3.4}$ &       88.3\,$\pm$\,11.9\\	 
2M1956$-$17 & M8 & L0 & M9 & L0\,$\pm$\,1 & 25.4$^{+3.3}_{-3.0}$ &	21.9\,$\pm$\,2.7\\	 
2M2243$-$59 & L1 & L1.5 & L0 & L0\,$\pm$\,1 & 29.4$^{+3.8}_{-3.4}$ &	34.0\,$\pm$\,4.2\\	 
2M2255$-$57 & L6 & L7.5 & - & L5.5\,$\pm$\,0.5 & 11.5$^{+1.3}_{-1.2}$ &  	85.8\,$\pm$\,9.0\\	 
2M2321$-13$ & L1 & L2 & L3 & L1\,$\pm$\,0.5 & 31.7$^{+2.1}_{-2.1}$   &	86.5\,$\pm$\,5.7\\    
\hline
\end{tabular}
\label{t2}
\end{minipage}
\end{table*}

Data reduction was undertaken within the {\sc iraf}\footnote{{\sc iraf} is distributed by the National
Optical Astronomy Observatories, which is operated by the Association of Universities for Research in Astronomy,
Inc. (AURA) under cooperative agreement with the National Science Foundation.} environment, using standard techniques for the
reduction of infrared spectroscopic data. As the data covers the region between the $J$ and $H$-bands, the master flatfields
contain regions where structure 
in the dispersion direction was present (owing to the water vapour column between the
dome and detector). These were interpolated using the {\it fixpix} routine. This means that in these regions there is
no correction for pixel-to-pixel variation, but only regions of strong telluric absorption are affected. 
A normalised flatfield was obtained by dividing the master flat by the same image smoothed by an 11$\times$11 pixel boxcar. 
All science and telluric standard frames were divided by the normalised flat. 

After flatfielding, science frames taken in each nod position were subtracted pairwise and spectra extracted using the {\it apall} routine. 
Optimal and non-optimal extractions were identical to within
typically $\la$\,a few\,\%. Arc spectra were extracted
using a profile of a relatively bright target star and calibrated with the {\it identify} and {\it reidentify} routines. 
The fit to the dispersion has typical 
RMS\,$\sim$\,0.1\,\AA. Individual spectra were only summed (using the {\it sarith} routine
with no pixel rejection) after wavelength calibration. Telluric correction was again performed using {\it sarith}; no attempts
were made to scale or shift the telluric spectrum to improve the correction. Hydrogen lines at 0.954, 1.005, 
1.092 and 1.282\,\micron~ were removed from the telluric spectra within the {\it splot} routine, before division. 
The telluric standard giving the best correction was normally the one closest in airmass and time but different standards 
from the same night were employed if they gave a better, less noisy, correction. In about 30\% of the cases, part of the 1.35\,\micron~
water band was interpolated over; in these cases there was too little flux in the region in either target or standard to provide a useful
correction. 
Finally the spectra were multiplied 
by a blackbody spectrum (F$_{\rm\lambda}$) with the T$_{\rm eff}$ of the chosen telluric standard star, and normalised to unity
at the highest point of the final spectrum.
 
\section{Spectral analyses}

In this Section, we detail the approaches made to analyse the target spectra and derive spectral types, distances and transverse velocities.
Additionally, alkali atomic line equivalent widths are measured in order to examine their behaviour with spectral type.
Moreover, spectral types from $J$-band spectral indices are measured and compared to those derived from direct comparison with
template objects.

\subsection{Direct comparison with templates}

Our primary method of spectral typing is direct comparison of target spectra with well-observed template objects
which have spectral types defined by near-infrared data. We have used the homogeneous dataset of \citet{cus05}
rebinned to the spectral resolution of our data, which is a factor of 2 smaller. These templates span the M and L
classes down to L8. Additionally, we have used our own observations of template objects, taken with the same instrumental setup
as the new objects. Best fits to the new data are shown in Fig.\,\ref{f3} where combinations of template spectra, with adjacent
spectral types ($\pm$\,1 subclass typically) from either our data or the \citet{cus05} data, or both, were created to
make a best fit. The method has the advantage of using data over the whole wavelength range observed and prevents
inaccuracies arising from e.g. differences in resolution between different datasets. 
For the remainder of this paper, we adopt these directly derived spectral types as the basis
for derivation of other quantities, notably the spectroscopic distance (see Table \ref{t2}). Fig.\,\ref{f3} shows all 24
spectra (solid line) with the best-fit template combination overplotted (dotted line). The derived spectral types and abbreviated
names are indicated. It may be seen that the overall quality of the fits is excellent. 
Note that for the three objects already observed spectroscopically \citep{lod05},
we obtain very similar types for two; for the third, 2M0123--42, \citet{lod05} quote L0.5, in contrast to our finding of M8\,$\pm$\,1.
We do not pursue the cause of this discrepancy here, but note that in a very recent publication \citep{rei06a}, a spectral type
of M8 is given.

\subsection{Spectral indices}

\begin{figure}
\includegraphics[bb=0 100 543 790,angle=270,width=9cm]{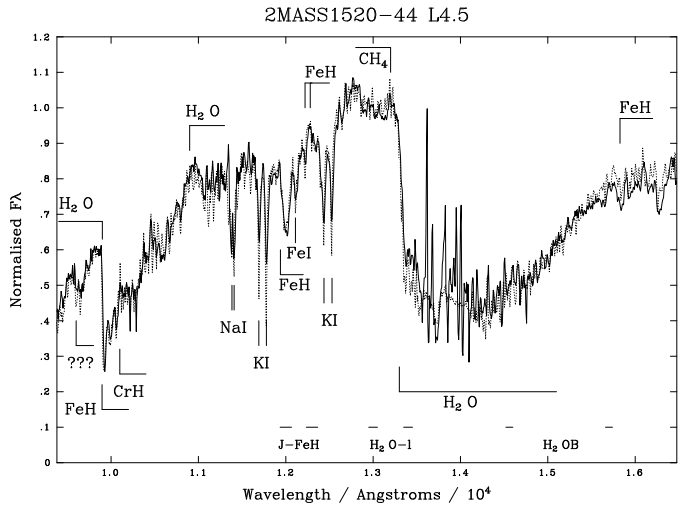}
 \caption{Close up of the spectrum of 2MASS1520$-44$, chosen for its high signal-to-noise ratio and late spectral type (L4.5).
The main features present in our wavelength range are labelled together with the locations and bandwidths of the spectral
indices we have used for spectral  analysis. The band at 9650\AA~is not yet identified with certainty but may be FeH; see the discussion in \citet{bur03}.}
\label{f4}
\end{figure}

We have used indices based on the H$_2$OB index of \citet{mcl03} together with the 
$J$-band FeH index of \citet{sle04} to predict spectral types for our objects. In addition we
use the H$_2$O-l index of \citet{sle04}, although this is inapplicable to $\sim$30\% of our objects
because of poor telluric correction, where no attempt has been made to derive the spectrum in the
deepest part of the water absorption feature around 1.35\,\micron.
The locations and bandwidths of these indices are shown in Fig.\,\ref{f4} together with the main atomic and molecular
features in an example mid-L object.

The results of
spectral typing by this method are given in Table \ref{t2} where applicable; in those cases where poor telluric correction precludes the use
of the H$_2$O-l index no value is given, and the FeH index is applicable only over spectral types earlier than $\sim$\,L3. 
Spectral types are based on the
relations provided by \citet{sle04} and \citet{mcl03} for the appropriate index.

For most objects, the spectral types derived from the indices agree well with those measured by direct comparison with templates.
In some cases there appear to be small differences, for example 2M0128$-$55, where both water indices predict a spectral type
$\sim$\,2 subclasses later. However the fit of template to observed data appears to be reasonable in this region, and in common with the other cases
where indices predict types either earlier or later than the template fit (not in a systematic fashion), these differences reflect
only scatter in the spectral type estimates; $\pm$\,1 subclass in general. The question of a
{\it systematic} difference in type suggested by the indices arises only in the cases of the two latest
type objects 2M1750$-$00 and 2M2255$-$57, where it is coupled with a poor fit of the template over part of the spectrum. These two cases
will be discussed further later in this paper.

\subsection{Spectroscopic distances and transverse velocities}

We have derived spectroscopic distances using our infrared spectral types and the M$_J$ vs. spectral type calibration of \cite{cru03}. Errors 
on the distances in Table \ref{t1} reflect the stated errors in spectral type propagated through the calibration.
Transverse velocities have been derived using the proper motions taken from SuperCOSMOS (Table \ref{t1}) and spectroscopic distances from Table \ref{t2}.
The results are also given in Table \ref{t2}. Errors on the transverse velocities are derived using only
the mean error on distance shown in Table \ref{t2}; with a few exceptions, errors on the proper motions are, relatively, smaller.
Although a few objects do show rather high transverse velocities, for example 2M0005$-$21, 2M0226$-63$, 2M1019$-$27, 2M1842$-$39 and 2M2255$-$57,
no object certainly has a velocity suggestive of halo kinematics. 
The last object on this list, 2M2255$-$57, stands out because it has much the
highest proper motion of this sample.  However, in common with three objects from \citet{ken04}, transverse
velocities of several tens of km\,s$^{-1}$ (rather than just a few tens) are observed and may be an indication that such objects do belong to a dynamically older
dwarf population. Indeed, according to \citet{bur04b}, a transverse velocity of 90\,km\,s$^{-1}$ may indicate halo kinematics when coupled
with a high radial velocity and high proper motion, as in the case of the two currently known L subdwarfs 2M0532+82 and 2M1626+39 
(\citealt{bur03}; \citealt{bur04b}). Our data are not of sufficiently high resolution to
infer radial velocities, and thus intrinsic space velocities, accurately. Further study of 2M2255$-$57, to derive its radial velocity
and therefore intrinsic space motion, is required to settle the matter. The same is true of the other objects
with rather high v$_{\rm trans}$ listed above, although only in the case of 2M2255$-$57 is there any possible {\it spectral} evidence 
for an unusual nature (see Sect. 5).

\subsection{Alkali line equivalent widths}

\begin{figure}
\includegraphics[width=9cm]{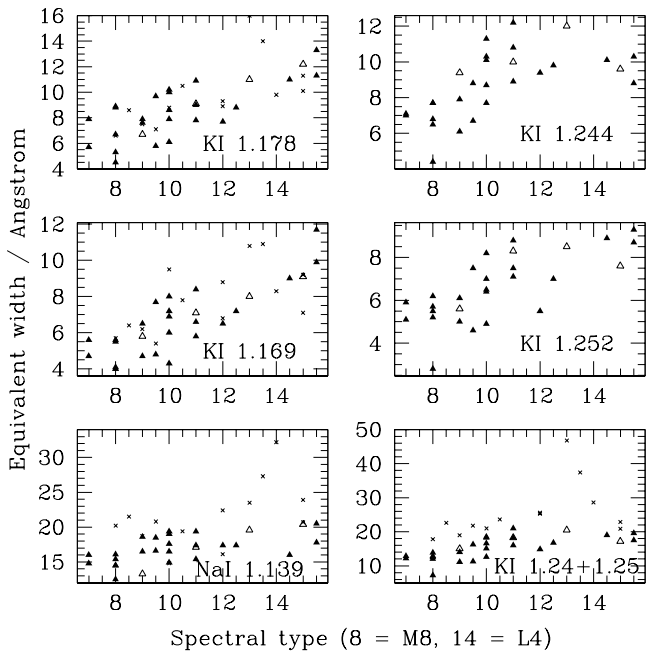}
 \caption{Equivalent width measurements for the 24 target objects (filled triangles) plus spectral standards observed with the same 
instrumental setup (open triangles) compared to equivalent width data from \citet{rei01} (crosses).}
\label{f5}
\end{figure}

In order to seek any objects with unusual spectral properties, for example older objects with lower metallicites and high gravities,
(which could be revealed as consistent with possibly halo-type kinematics, as discussed above) 
we have plotted in Fig.\,\ref{f5} equivalent widths
of the strong K\,{\sc i} and Na\,{\sc i} lines seen in our spectra, against spectral type. Also plotted as crosses are equivalent width
data for a set of well-known ultracool dwarfs from \citet{rei01}. Equivalent widths
were measured relative to the local pseudocontinuum using the {\sc iraf} {\it splot} routine. Errors
are $\sim$\,$\pm$ 1--2 \AA~ and larger for the Na\,{\sc i} line which is affected by telluric
correction uncertainties and clearly shows a greater scatter. However, as is clear from \citet{bur03}, L subdwarfs
show a rich $J$-band spectrum with atomic line strengths very similar to ``normal'' field L dwarfs although the
lines are broader in the high gravity objects. Alkali lines are also strong in extreme late M subdwarfs
\citep{bur06}. Hence atomic line strengths alone are not diagnostic of low metallicity; but also we see no evidence for unusually broad atomic lines in any of our spectra.
We can though rule out the existence in our sample of young, low-gravity objects with very weak atomic lines and enhanced VO,
as seen in the young field dwarf 2M J0141$-$46 \citep{kir06}. Moreover, we observe excellent agreement between the K\,{\sc i}
1.169 and 1.178\,\micron~ equivalent widths of \citet{rei01} and our measurements. Hence, within the limitations of
the data, we can conclude that none of our sample spectra appear unusual in this respect.  
As noted, the Na\,{\sc i} line is affected by telluric absorption and while it appears that
the equivalent widths of \citet{rei01} are systematically larger than our measurements, this is most likely the effect of the
resulting large measurement errors. In the case of the sum of the K\,{\sc i} 1.244 and 1.252\,\micron~lines in our objects, plotted
in the lower right panel of Fig.\,\ref{f5} for comparison with \citet{rei01}, their measurements again appear
systematically larger than ours. However it is likely the \citet{rei01} values are measured as a blend, 
whereas ours are the sum of the individual line measurements. As can be seen from Figs.\,\ref{f3}
and \ref{f4}, the lines of this doublet are more blended in our spectra than those of the 1.17\,\micron~pair.    

Thus, taking into account all these considerations, it is clear that the equivalent widths of the
atomic lines in our new objects do not appear to behave differently from that predicted by their spectral type.
Therefore, the analysis of alkali lines offers no evidence for the presence of low metallicity and/or high gravity (subdwarfs)
or low gravity (young objects) in the new sample of 24 ultracool dwarfs presented here.

\section{Discussion}

In the analysis of $JH$ band spectra of twenty four ultracool dwarfs presented above, of which twenty-one are hitherto
undescribed in the literature, spectral types ranging from M7 to L5.5 are found by direct comparison to template
objects taken from the homogeneous dataset of \citet{cus05} together with five well described standards we have observed
with the same instrumental setup. In general, we find no evidence that any of the 21 new objects have peculiar metallicities
and/or kinematics. In this Section, we select individual objects from the sample for further discussion. 

In particular, 
two L5-L6 dwarfs, 2M1750$-$00 and 2M2255$-57$, stand out in the sample, with
spectroscopic distances near 10\,pc. (2M1520$-$44, which we also place near 10\,pc in Table \ref{t2}, will be discussed in Sect. 6).
2M2255$-$57 has a large proper motion and a relatively high
transverse velocity, 86 km\,s$^{-1}$, while 2M1750$-$00 does not stand out from the sample in either
respect. It is, however, apparently one of the brighter sample objects ($J$\,=\,13.29) and clearly nearby, $\sim$\,8\,pc
by our analysis.

For 2M1750$-$00 and 2M2255$-$57, both H$_2$O-band indices (Table \ref{t2}) predict later L spectral
types, nearer L7--L8. Indeed, Fig.\,\ref{f3} shows that the strong H$_2$O absorption feature is not readily fit by a combination of
L5 and L6 spectral types. The reason for this is not clear. In the case of 2M2255$-$57, it has already been
suggested that there is marginal kinematic evidence for membership of a dynamically old population. It is conceivable, then, that
the apparently overly strong H$_2$O could arise from its enhanced gravity, as is evident in the two known
halo sudwarfs 2M0532+32 and 2M1626+39 (\citealt{bur03}; \citealt{bur04b}). In this scenario, metal hydride bands would also be enhanced,
leading to an under-estimation of the spectral type from the $J$-band, where the FeH bands are well fit by the L5-L6 templates.
However there is no kinematic evidence for this hypothesis in the case of the other L5.5 object 2M1750$-$00. More importantly, for neither object 
is there photometric evidence
of collision-induced H$_2$ absorption which would suggest a high-gravity atmosphere. This is known to result in unusually
blue near-infrared colours ($J-K$\,$\sim$\,0.3; \citealt{bur04b}) whereas both objects here have perfectly normal $J-K$ colours, $\sim$\,1.5.
Therefore, we can probably cast doubt on the high-gravity hypothesis for both these L5.5 dwarfs, in spite of the marginal kinematic evidence for 2M2255$-$57.

Another possibility is that both these objects are in fact unresolved binaries, with later L-type companions with redder colours contributing
relatively more to the $H$-band than the $J$-band. This would explain the fact that L5-L6 fits the $J$-band well, especially FeH. With FeH
weakening towards later L types, these features in the $J$-band would be unaffected by a later companion, which could however explain the observed, deeper
$H$-band H$_2$O band, since the strength of the water absorption continues to increase throughout the later L types and into the T dwarf
regime. There is no evidence for asymmetry in our acquisition images. It is clear, though, these two objects are therefore of special interest for
follow-up high resolution (AO or space-based) imaging in order to check for the existence of later-type companions. In this context, 
we note the recent discovery of the L/T binary 2MASSJ\,22521073$-$1730134 \citep{rei06b}.

Finally, a general point about the whole sample; 
we note that although the survey is biased towards bright objects (with optical detections) it is possible that the use
of the reduced proper motion may have led to a very bright, high proper motion object being missed. For example, for an object
with a proper motion of 500\,mas\,yr$^{-1}$ to lie towards the left of the distribution of known ultracool dwarfs in Fig.\ref{f2} (for
example at H$_{\rm J}$\,=\,23) it would then be a bright object with $J$\,=\,9.5. For an M6 dwarf ($M_J$\,$\sim$\,10) this corresponds to
d\,$\sim$\,8\,pc and for an L8 dwarf ($M_J$\,$\sim$\,14.5), d\,$\sim$\,1\,pc. It is extremely unlikely that any such object would have
been overlooked by a proper-motion only survey. However, low proper motion (low velocity) dwarfs certainly remain to be found in the
remainder of the photometrically selected candidates with H$_{\rm J}$\,$<$\,25 which we have not yet observed spectroscopically.     

\section{Discovery of a resolved binary L dwarf: 2MASSJ\,15200224$-$4422419}

\begin{figure}
\includegraphics[angle=270,width=8cm]{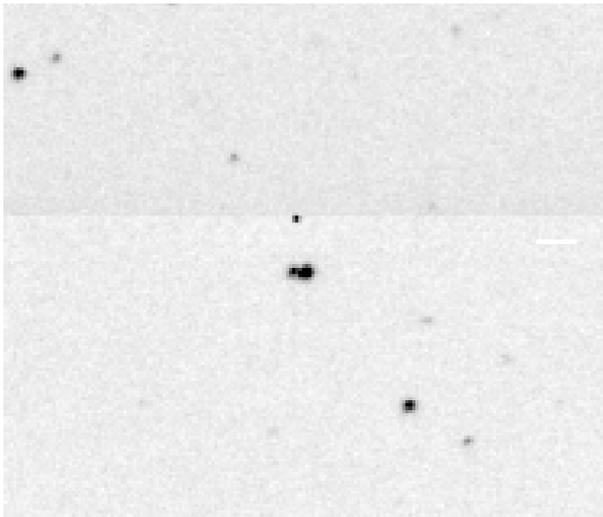}
 \caption{SOFI acquisition image for 2M1520$-$44 in the $J$-band. The separation is $\sim$1\arcsec.}
\label{f6}
\end{figure}

\begin{figure}
\includegraphics[angle=270,width=9cm]{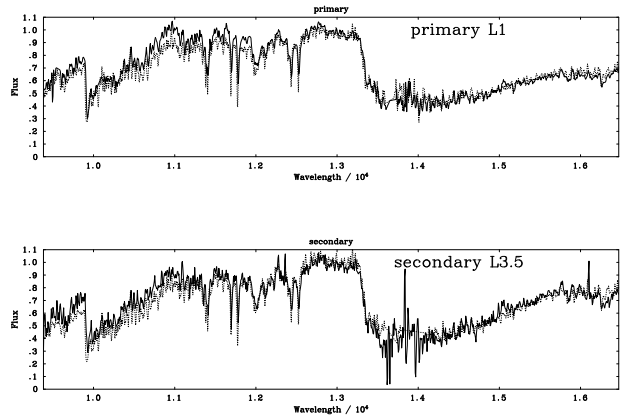}
 \caption{Spectral types for the two components by direct comparison to the \citet{cus05} objects 2M1439+19 (L1) and 2M0036+18 (L3.5).}
\label{f7}
\end{figure}

A spectral type of L4.5 and a spectroscopic distance near 10\,pc were initially derived for 2M1520$-$44. However, SOFI acquisition 
imaging clearly resolved this object as a binary during the 2006 April 4-7 run. Therefore,
supplementary spectroscopic data for 2M1520$-$44 were obtained on 2006 April 11 where the slit was aligned across both components.
The instrumental setup was exactly as described in Sect. 3, and data reduction was performed in a similar way. The clear binary
nature of this L dwarf is apparent in the acquisition image shown in Fig.\,\ref{f6}. However, poor seeing meant that the
two adjacent stellar profiles were blended. Hence it is likely that flux from one object contaminates the other, and 
the extracted spectra are still composite to some extent. However, using the {\sc starlink} package {\it figaro}, we were able to extract
a spectrum for both the primary and secondary component separately. Fig.\,\ref{f7} shows the two separate spectra for which we estimate $\sim$\,L1
and $\sim$\,L3.5 for the primary and secondary components, respectively, by direct comparison to two \citet{cus05} templates.

Furthermore, we employed the H$_2$OA and H$_2$OB indices of \citet{mcl03} to estimate spectral types of L2.5 and L1.5 for the primary,
and L4 and L4.5 for the secondary, consistent with the results shown in Fig.\,\ref{f7} to within $\pm$\,1 subclass. Hence the derivation of
L4.5 for the composite spectrum (Fig.\,\ref{f3} and Table \ref{t2}) seems strange. However, it is possible that a misalignment of the slit
might have caused the earlier observation to be of the secondary alone, without much contribution from the primary. 

We were unable to perform photometric measurements accurately on the acquisition image owing to a lack of calibration data. However,
the strength of the stellar profiles in the two-dimensional spectra taken on 2006 April 11 are approximately in the ratio 1:2.5, i.e.
the primary is 2.5 times brighter; one magnitude. Adopting
spectral types L2 and L4 for the two components, we can derive M$_J$\,=\,12.3 and 13.1 from the spectral type vs. absolute $J$ magnitude
relation of \citet{cru03}. This difference of $\sim$1\,mag is exactly what is observed, implying that the data are quite consistent
with both components lying at the same distance, and being physically associated. Indeed, the flux ratio of 1:2.5 observed, together
with the 2MASS $J$ magnitude, allow us to derive $J$\,=\,14.57 for the secondary and $J$\,=\,13.60 for the primary, and distances of 19.7 and 18.2\,pc
respectively. These two values are consistent within the likely errors. Suitable data are not available for a common proper motion test; the object
is too far south to be in the SDSS data, for example. We conclude that 2M1520$-$44 is very likely a binary, and note that the secondary has
a strong likelihood of being substellar.

\section{Conclusions}

Using a combination of optical and near-infrared survey data from SuperCOSMOS and 2MASS, twenty-one hitherto
undescribed sources are shown to be ultracool dwarfs with spectral types in the range M7 to L5.5 using low resolution
spectroscopy from NTT/SOFI. The sample objects have $JHK$ magnitudes and colours comparable to
known late M and L dwarfs and are bright enough to have optical detections in digitised plate data. They have proper motions
typically a few $\times$10$^2$ mas\,yr$^{-1}$.
Spectroscopic distances are derived, typically within 30\,pc, showing these objects to be important additions to our knowledge
of low-mass, low-luminosity objects in the Solar neighbourhood.

Two objects in particular, 2M1750$-$00 and 2M2255$-$57, are of particular interest, with spectral types L5-L6 and spectroscopic
distances near 10\,pc. The latter has a high proper motion, 1.5\arcsec\,yr$^{-1}$.
These two objects are likely brown dwarfs. In common with the rest of the sample, they are of great
interest for further study using AO, HST, or methane imaging, to uncover lower-mass, later type companions. There is circumstantial
evidence for the existence of such companions from the spectroscopic data presented here. Furthermore, these two objects in particular
merit a parallax determination. 

We have found a further object, 2M1520$-$44, to be a resolved L dwarf binary consisting of L2 and L4 components, lying $\sim$\,19\,pc distant. Further study
should reveal this source as a common proper motion pair. As such, it may prove a useful test-bed for evolutionary models of very low-mass stars and 
brown dwarfs.

\section*{Acknowledgments}

TRK acknowledges financial support from PPARC. He would also like to
thank the staff at ESO La Silla and especially Valentin D. Ivanov
for his support of this project. 
This research has made use of data obtained from the SuperCOSMOS Science Archive, 
prepared and hosted by the Wide Field Astronomy Unit, Institute for Astronomy, 
University of Edinburgh, which is funded by the UK Particle Physics and Astronomy 
Research Council. This research has benefitted from the M, L, and T dwarf compendium housed at 
DwarfArchives.org and 
maintained by Chris Gelino, Davy Kirkpatrick, and Adam Burgasser. This research has made use of the SIMBAD database,
operated at CDS, Strasbourg, France.

\label{lastpage}
\end{document}